\documentclass{article}
\usepackage[utf8]{inputenc}
\usepackage{amsmath}
\usepackage{graphicx}
\usepackage{geometry}
\usepackage{graphicx}
\usepackage{caption}
\usepackage{subcaption}
\usepackage{natbib}
\usepackage{comment}
\usepackage{setspace}
\usepackage{lineno}
\title{The Temporal Onset of Habitability For Earth-Like Planets}
\author{Johnny Seales and Adrian Lenardic}

\begin{document}

\maketitle

\textbf{The ability of a planet to maintain surface water, key to life as we know it, depends on solar and planetary energy. As a star ages, it delivers more energy to a planet. As a planet ages it produces less internal heat, which leads to cooling. For the Earth, interior cooling connects to plate tectonics - the surface manifestation of convection within the Earth's interior. This process cycles volatiles (CO2 and water) between surface and interior reservoirs, which affects planetary climate. Cycling rates depend on the efficiency of plate tectonic cooling. That efficiency remains debated and multiple hypotheses have been put forth. Geological proxy data allow us to validate these hypotheses accounting for model and data uncertainty. Multiple models pass the validation test. Those models define a distribution for terrestrial exoplanets akin to Earth, accounting for variations in tectonic efficiency. Feeding this distribution into climate models indicates that the time at which habitable conditions are established can vary by billions of years. Planets of the same absolute age and orbital distance can reside and not reside within the classic habitable zone due to differences in plate tectonic cooling efficiencies. The full model population allows a probability distribution to be constructed for the the time at which habitable conditions are established. The distribution indicates that Earth-like exoplanets, of the same age, can be at different evolutionary stages. It also indicates that planets around stars whose early evolution is unfavorable for life can become habitable later in their energetic histories.}

The cycling of volatiles between a planet's interior and surface depends on volcanism and tectonics, which are driven by a planets internal energy. This connects volatile cycling to a planet's internal cooling history \citep[e.g.,][]{McGovern1989}. For Earth, dated rock samples provide thermal history constraints \citep[][]{Herzberg2010a, Condie2016a, Ganne2017}. However, model and data uncertainties allow multiple cooling paths to remain viable, even after taking an Occam's razor approach that the principal cooling mechanism for the modern Earth, plate tectonics, has been operative over its geologic history \citep[][]{Seales2020}. 

\begin{figure}[htb!]
\centering
  \includegraphics[width=.5\linewidth]{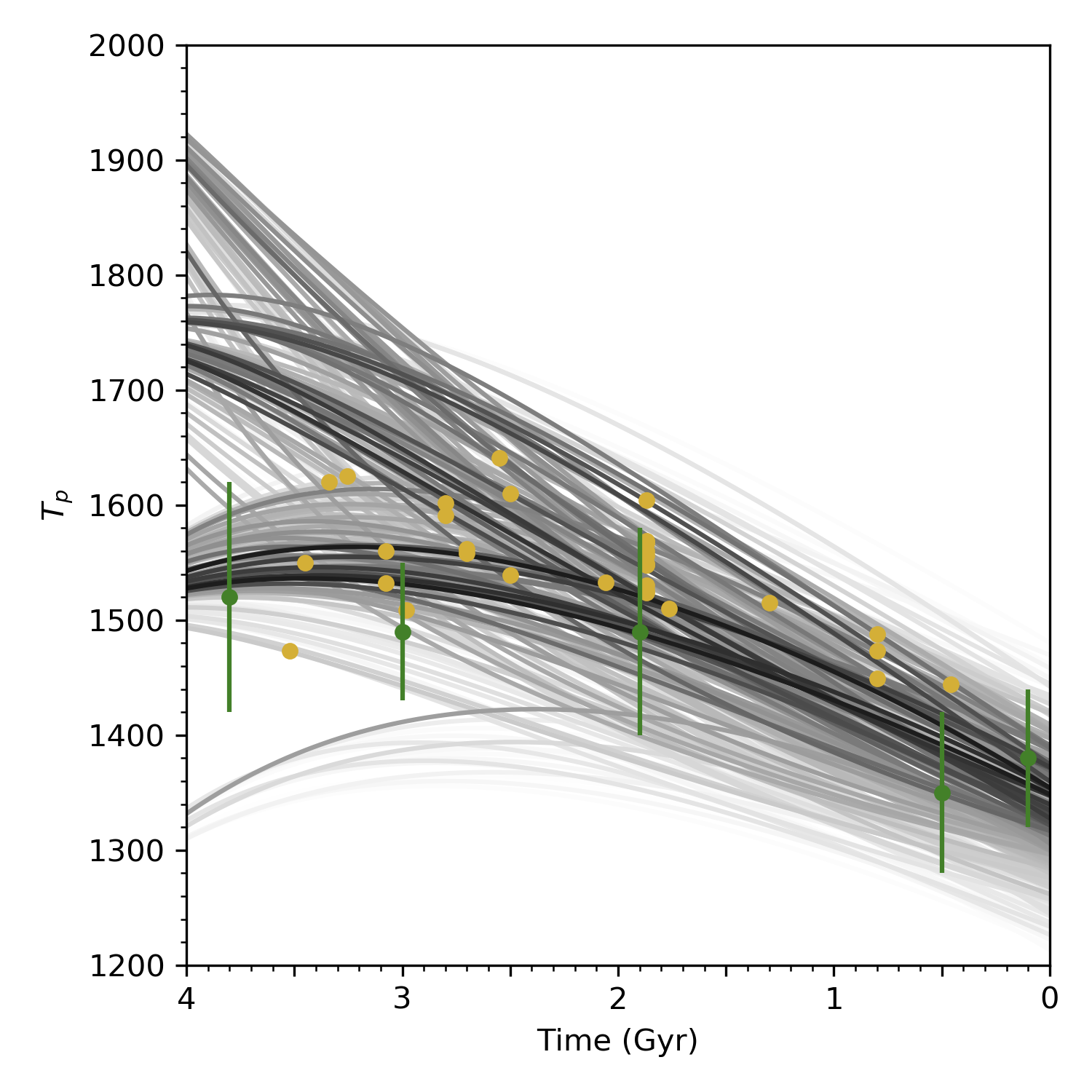}
\caption{Black lines show the range of thermal histories, each computed with different initial conditions, parameters and underlying physical models, that can match Earth geological proxy data and are used to calculate surface temperature histories. Darker lines indicate higher probability for particular model ensembles to match data constraints.}
\label{ECD}
\end{figure}

\citet{Seales2020} took an agnostic modeling approach that evaluated a range of hypotheses for plate tectonic cooling efficiency. They calculated more than a million model cooling paths and evaluated probabilistically how well each matched data constraints (Figure \ref{ECD}). Figure \ref{ECD} shows paths that matched constraints within uncertainty. Each line represents the average of an ensemble - a set of paths, for equivalent model parameter values and initial conditions, subjected to a perturbed physics analysis to determine the structural uncertainty of models \citep[][]{Seales2019}. The 2-sigma uncertainty envelopes used for assessing the probability of an ensemble matching data constraints is left off of Figure \ref{ECD} for clarity. We retain a visual measure of uncertainty by shading lines according to the probability that an ensemble matched data constraints. The Earth followed a single cooling path, but data and model uncertainty preclude uniquely determining that path. The Earth may have followed the most probable path, but we have no reason for assuming a single planet should fall in the peak of a probability distribution. More key for the this paper, we have no \emph{a priori} reason for assuming that the tectonic efficiency of the Earth cannot differ from other terrestrial planets that share Earth characteristics. Thus, all the paths remain viable. 

\begin{figure}[htb!]
\centering
  \includegraphics[width=\linewidth]{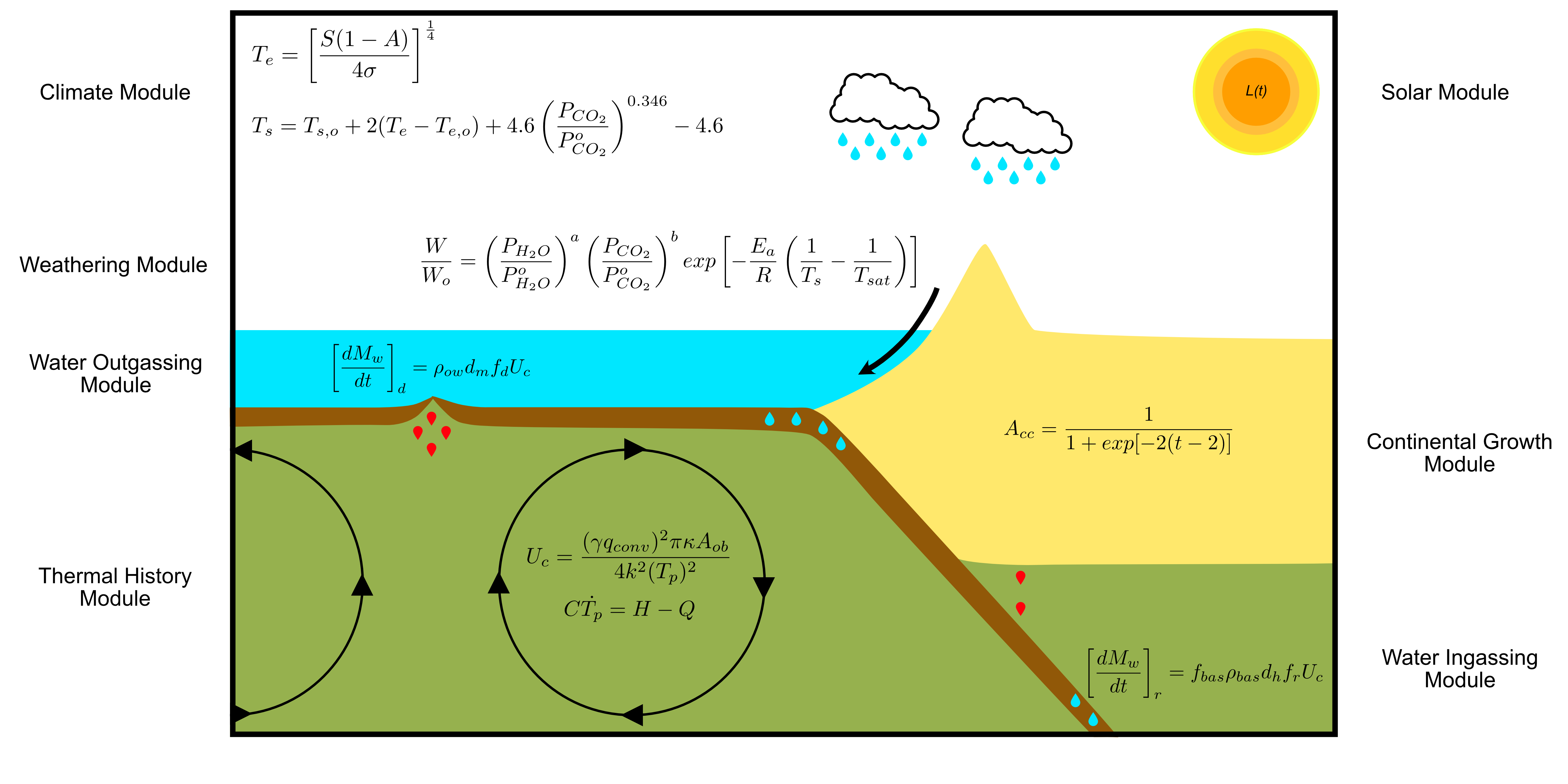}
\caption{Cartoon schematic depicting the different modules. Details of each module can be found in the appendix.}
\label{fig:Cartoon}
\end{figure}

The subset of viable thermal history paths provides the base of our study. Coupling these solid planet models to volatile cycling and climate models (Figure \ref{fig:Cartoon} provides a method for evaluating how variable plate tectonic cooling efficiencies impact surface temperatures. Using these, we can estimate the time window over which the surface becomes habitable for Earth-like planets. Our definition of Earth-like in this context is more specific than the way the term is generally used (i.e., terrestrial planets of comparable size and mass). Our Earth-like distribution only includes models that can match data constraints on Earth cooling. This tightens the distribution span. With knowledge of our results, we can say that this will already lead to a large temporal span for surface temperature evolution. Expanding the approach to consider tectonic modes other than plate tectonics and/or planets that differ from the Earth in other ways (e.g. size, composition) would allow for cooling paths that are not consistent with Earth data. This would amplify the conclusion that the timing at which temperate surface conditions are established can vary significantly for planets of the same age and same orbital distance from stars with comparable evolutions. 

To maintain focus on the role of tectonic cooling efficiency, free of added novelty, we adopt volatile cycling and climate modules previously employed in habitability research \citep[][]{Walker1981a,Foley2016,Mello2020}. Figure \ref{fig:Cartoon} shows the different modules involved in our coupled model. The full model is described in Supplementary Materials. Different tectonic efficiencies result in different volatile cycling rates. Volatiles (water and CO2) are degassed at spreading centers while subducting plates carry them back into the Earth's interior \citep[][]{Cowan2014a}. The balance of CO2 in the atmosphere is modeled through a silicate-weathering feedback \citep{Walker1981a}. The transfer of carbon from the interior and weathering rates, which depend on topography, are parameterized with plate velocities \citep{Walker1981a}. This connects greenhouse gas levels in the atmosphere to the internal cooling rate of a planet. Silicate-weathering operates in tandem with evolving solar flux \citep{Rushby2013}. This allows surface temperatures to depend on the co-evolution of solar and internal planetary energy. Climate evolution also depends on solar distance and albedo which, for the purposes of our study, remain fixed over model evolution time but can be varied from model to model. 

\begin{figure}[h!]
  \centering
    \begin{subfigure}[b]{0.35\textwidth}
      \includegraphics[width=\linewidth]{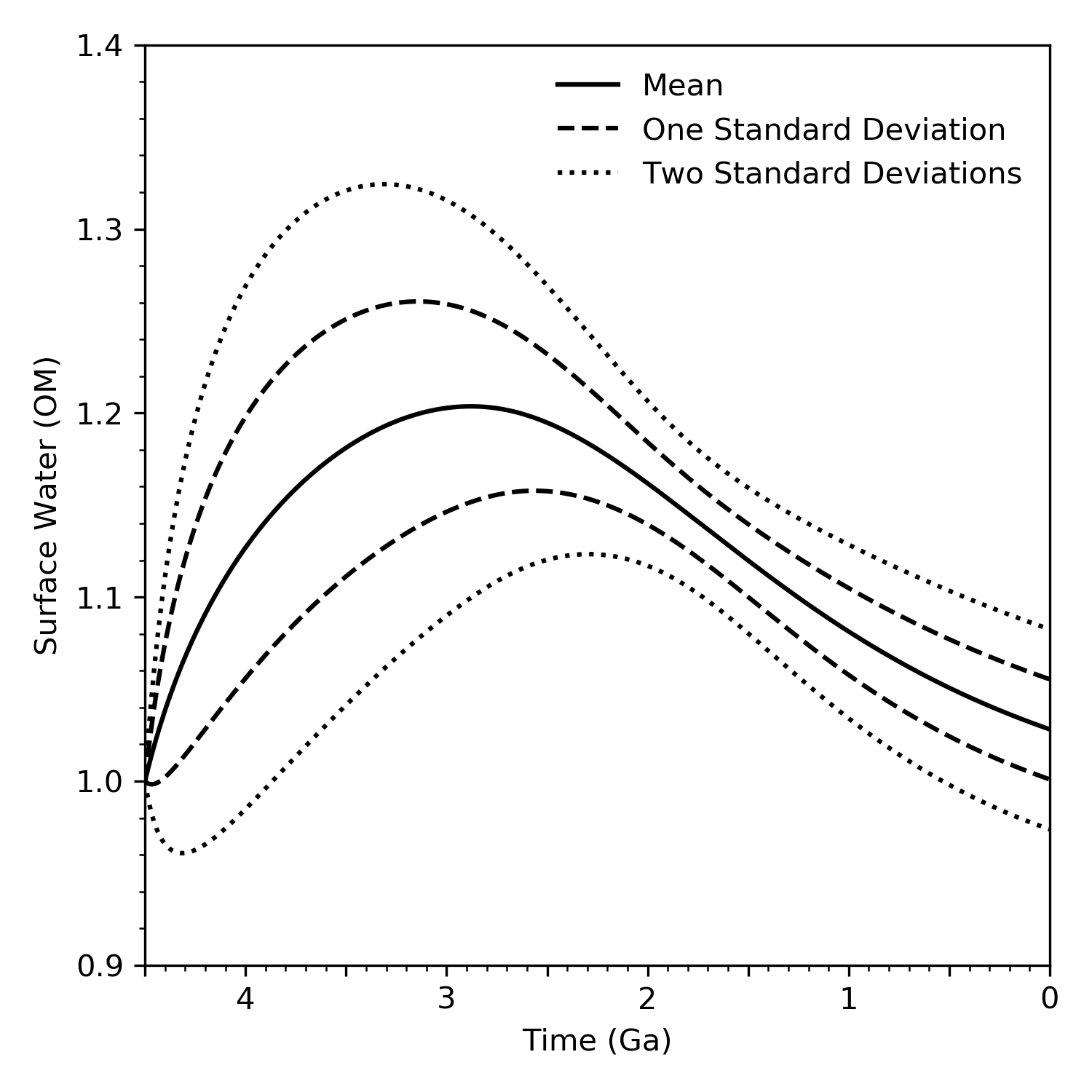}
      \caption{}
      \label{fig:Earth_water}
    \end{subfigure}%
    \begin{subfigure}[b]{0.35\textwidth}
      \includegraphics[width=\linewidth]{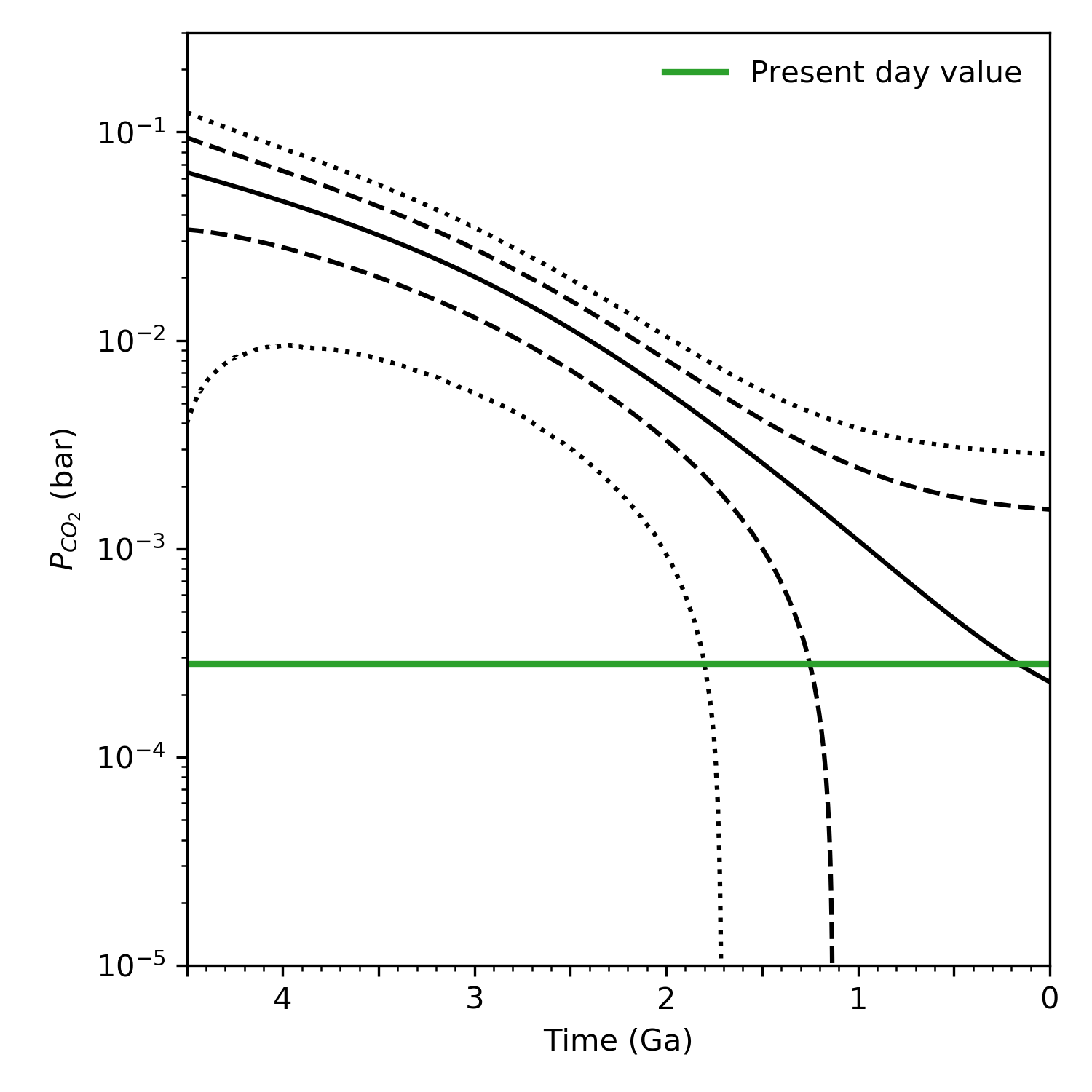}
      \caption{}
      \label{fig:Earth_PCO2}
    \end{subfigure}%
    
    \medskip
    \begin{subfigure}[b]{0.35\textwidth}
      \includegraphics[width=\linewidth]{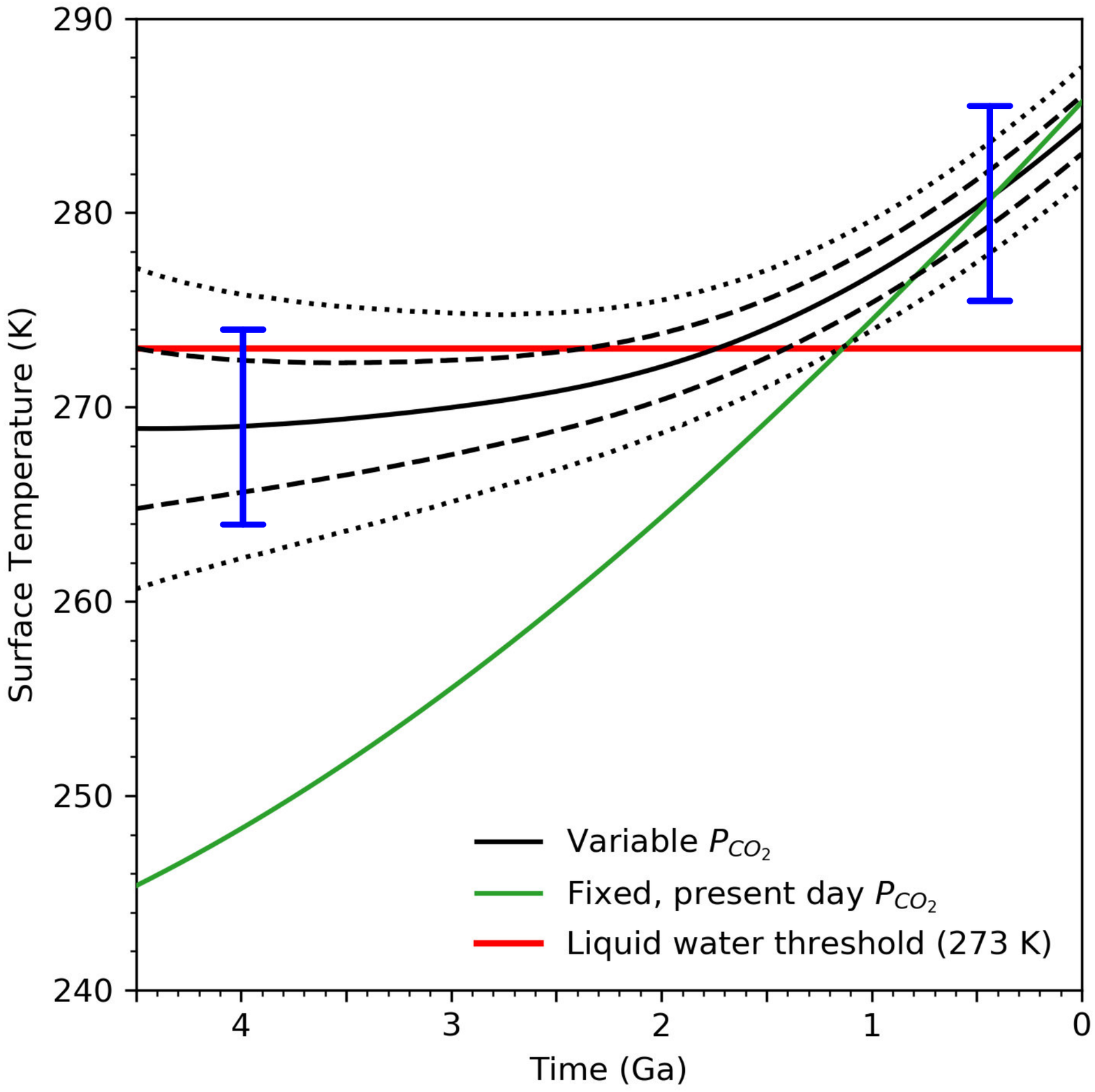}
      \caption{}
      \label{fig:Earth_Ts}
    \end{subfigure}%
    \begin{subfigure}[b]{0.35\textwidth}
      \includegraphics[width=\linewidth]{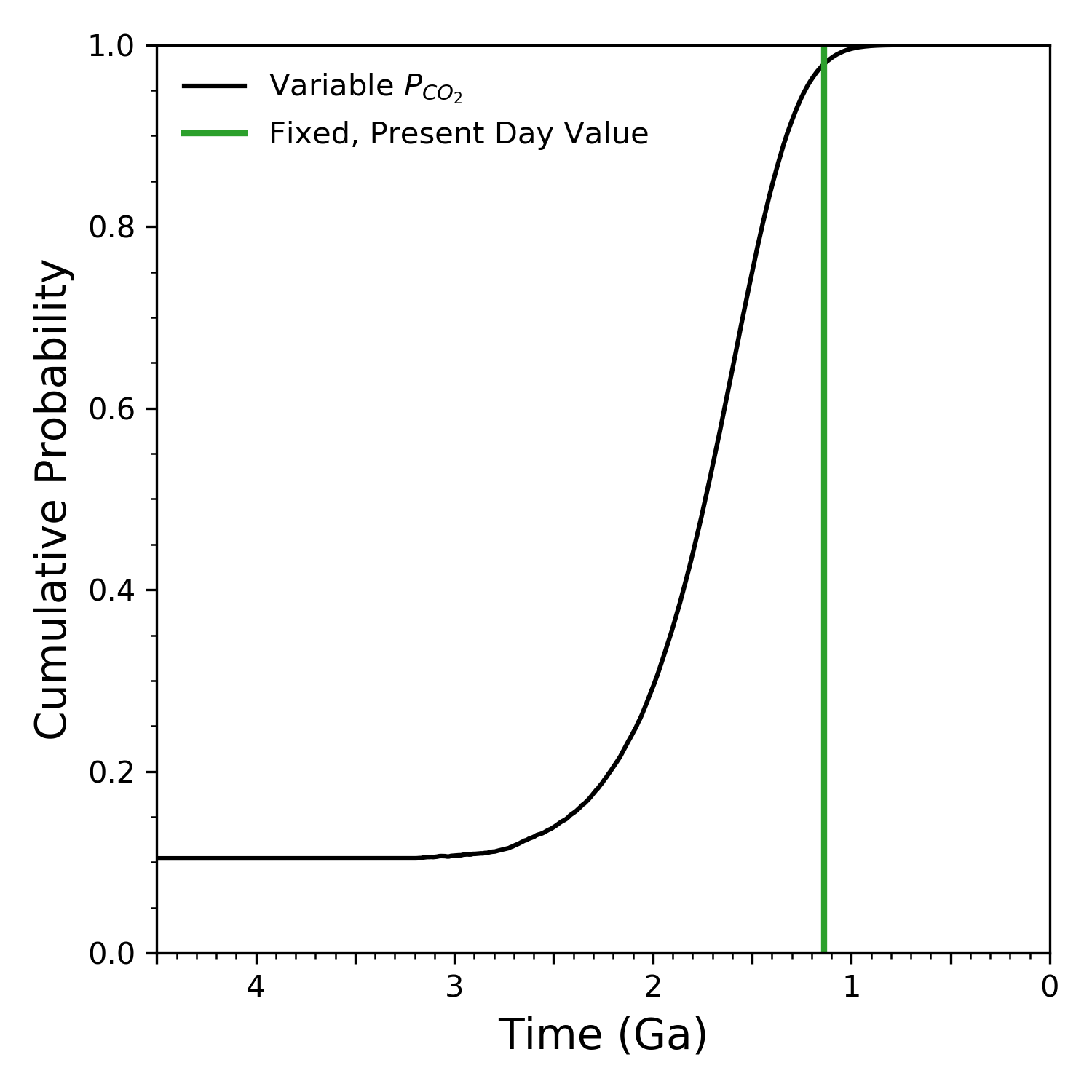}
      \caption{}
      \label{fig:Earth_EHZ}
    \end{subfigure}
\caption{Uncertainty in surface water (a) atmospheric $CO_2$ (b) and surface temperature (c) for Earth constrained thermal histories. The blue error bars represent the magnitude of fluctuations about the mean. Cumulative probability for the timing of temperatures rising above freezing (d) relative to the predicted entry time for present day $P_{CO_2}$ and one ocean mass of water (green line).}
\label{fig:Earth_outputs}
\end{figure}

Figure \ref{fig:Earth_outputs} shows surface water (panel a), atmospheric $P_{CO_2}$ (panel b), and mean surface temperature (panel c) evolutions, plotted as uncertainty windows, for models at Earth orbital distance and current global albedo. Also plotted is how surface temperature would have varied for $P_{CO_2}$ fixed at present day levels (green line). The difference between that trend and other model paths shows that volatile cycling works to offset a faint young sun but to variable degrees depending on plate tectonic cooling efficiency. 

Traditionally, thermal history models track how mean tectonic plate speeds, and associated internal cooling, evolve in response to decaying internal heat sources \citep{Davies1980, Schubert1980}. This is referred to as a secular trend. However, plate tectonics allows for variations in plate speeds that occur over a shorter time scale than that associated with secular decay of heat sources \citep{Lenardic2016c}. Our approach allows us to model temporal fluctuations about mean values \citep{Seales2019}. Those tectonic fluctuations, when fed into climate models, lead to surface temperature fluctuations on the order of 10 degrees over a hundred million year time scale. This is consistent with constraints for the Earth's paleo climate and with a study that explicitly modeled climate variations associated with super-continent aggregation and dispersal \citep{Jellinek2020}. Figure \ref{fig:Earth_outputs} shows the magnitude of temperature fluctuations on one of the secular trends for mean surface temperature (all of the evolutions allow for similar fluctuations about mean trends).

Figure \ref{fig:Earth_outputs} shows  range of potential behavior that can be grouped into sub-suites. For one sub-suite, mean surface temperature paths maintain conditions favorable for liquid water over full model time. For those cases, elevated tectonic cooling rates, and enhanced C02 cycling, offset the faint young sun. A second sub-suite has mean temperatures below the freezing point of water with tectonic fluctuations allowing for periods of warmer conditions. A third sub-suite had high enough CO2 degassing to offset the faint young sun early in a planets evolution but as the planet lost internal energy and tectonic cooling rates decreased, surface temperatures dropped below freezing. When the solar flux grew larger, the mean surface temperature returned above freezing. A final suite allowed delayed entry into the habitable zone that varied by roughly a billion years. 

Figure \ref{fig:Earth_EHZ} shows the cumulative probability that Earth-like planets entered the habitable zone at any given time. We calculated it by determining the percentage of paths that remained above freezing (red line) at anytime (following the classic definition of the habitable zone, we track this in terms of mean surface temperatures). For those that began with liquid water but subsequently froze, we used the most recent thawing to define a cumulative probability. For reference, Figure \ref{fig:Earth_EHZ} shows the Earth would have entered the habitable zone approximately one billion years ago if it had present day $P_{CO_2}$ values throughout its history. Increased volcanism, associated with higher temperatures in the Earth's past, produced elevated $P_{CO_2}$ levels. As such, models could entered the habitable zone earlier than would be the case without volatile cycling. None the less, variations in tectonic cooling allowed the entry time to vary by roughly 3 billion years for the full distribution of Earth-like planets. Evidence of liquid water on Earth more than four billion years ago \citep{Mojzsis2001} suggests that the Earth may fall in the tail rather than the center of the probability distribution for Earth-like planets or that its early history may have oscillated into and out of the classic habitable zone. 

\begin{figure}[h!]
  \centering
    \begin{subfigure}[b]{0.35\textwidth}
      \includegraphics[width=\linewidth]{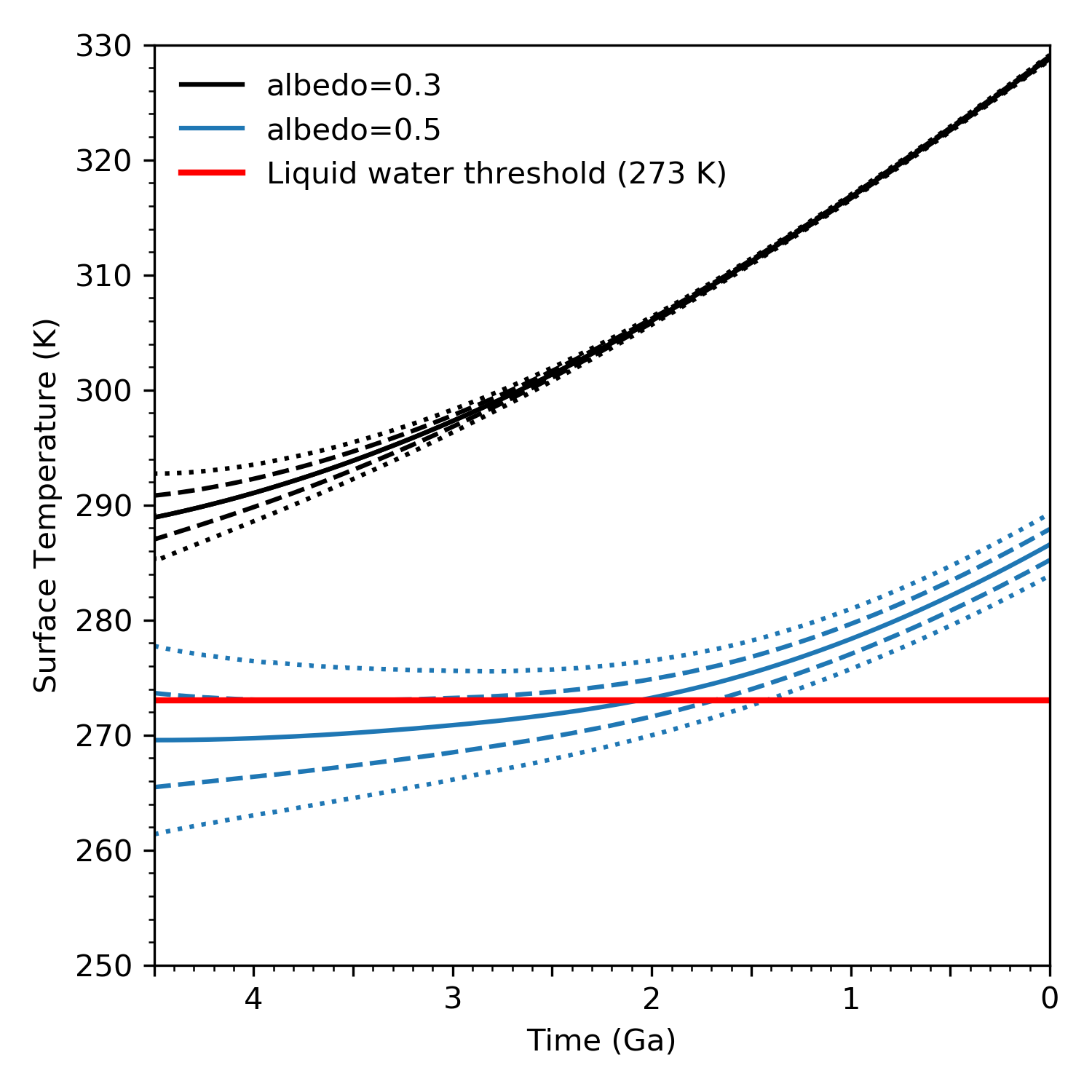}
      \caption{}
      \label{fig:VenusTs}
    \end{subfigure}%
    \begin{subfigure}[b]{0.35\textwidth}
      \includegraphics[width=\linewidth]{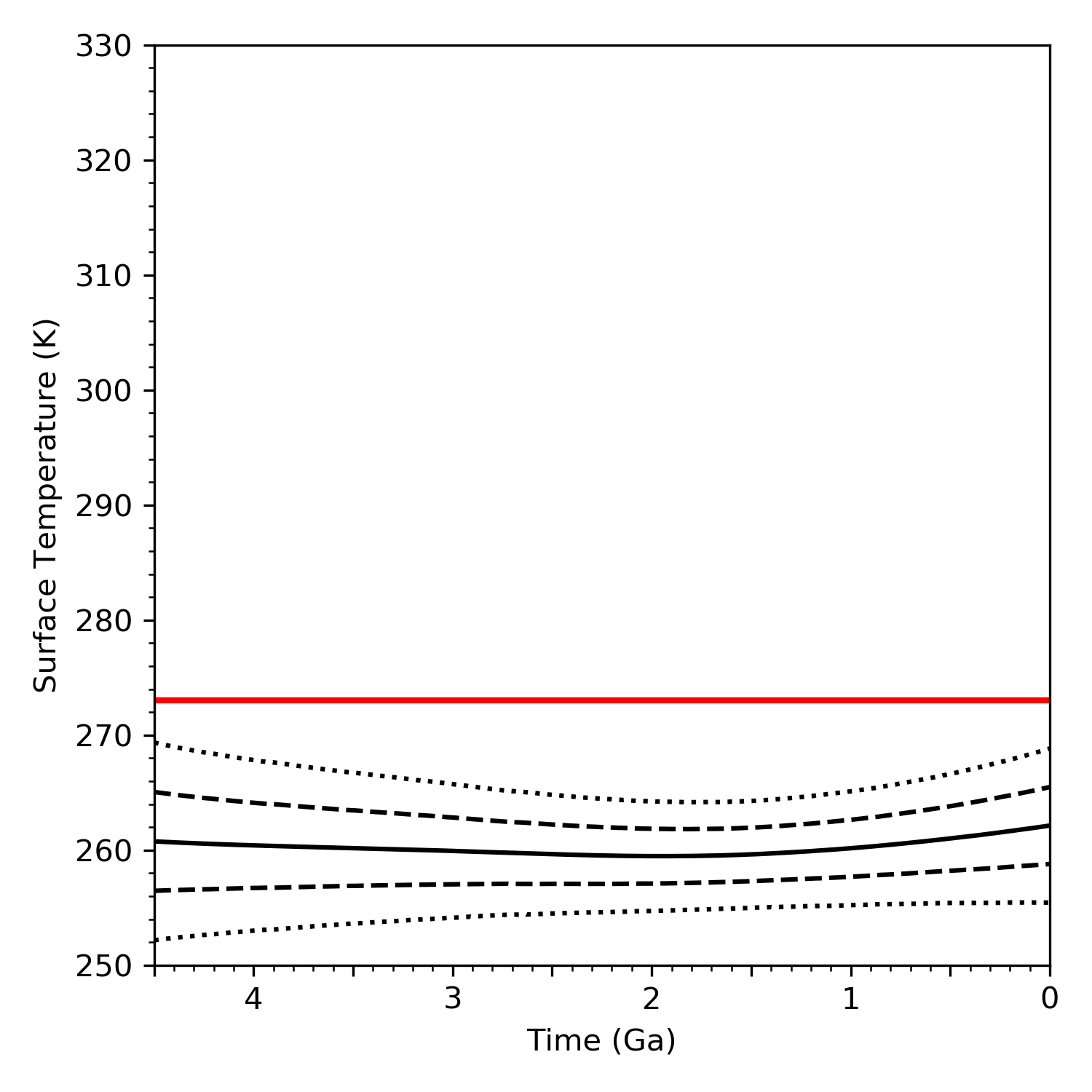}
      \caption{}
      \label{fig:MarsTs}
    \end{subfigure}%
    
    \medskip
    \begin{subfigure}[b]{0.35\textwidth}
      \includegraphics[width=\linewidth]{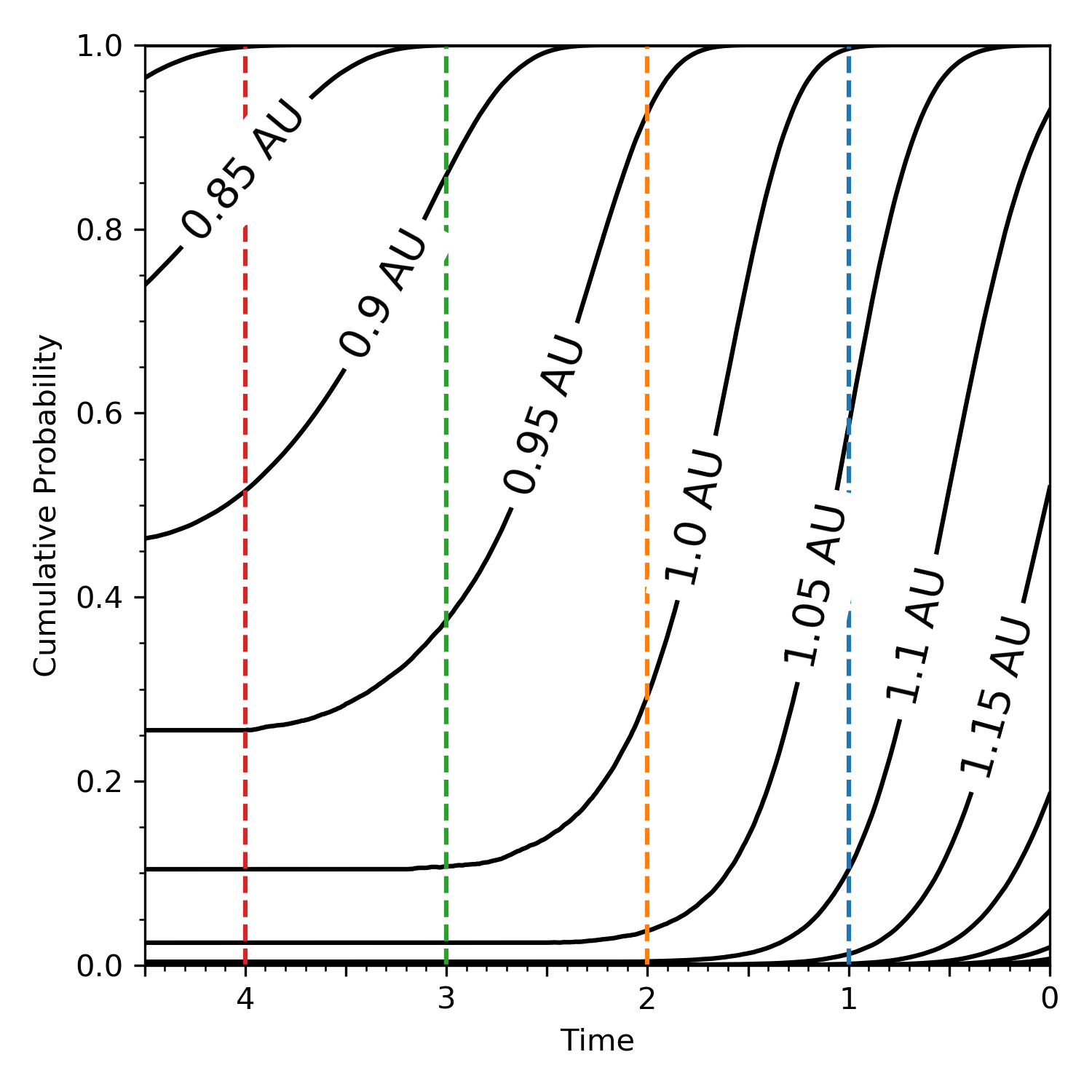}
      \caption{}
      \label{fig:DCDF}
    \end{subfigure}%
    \begin{subfigure}[b]{0.35\textwidth}
      \includegraphics[width=\linewidth]{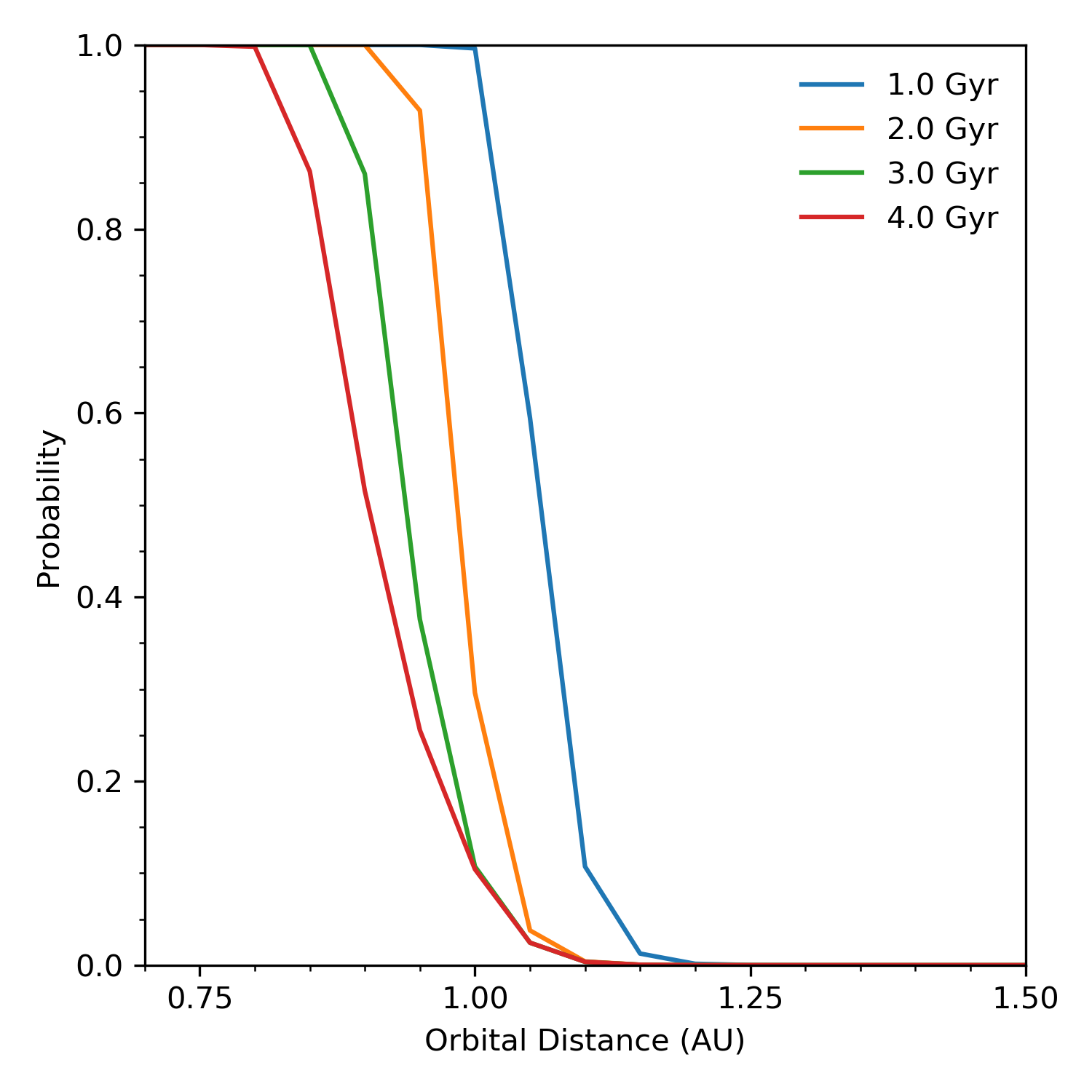}
      \caption{}
      \label{fig:DEHZ}
    \end{subfigure}%
\caption{Uncertainty in the surface temperature at a Venusian (a) and Martian (b) orbital distance. We fixed abledo at 0.3 for both and tested an albedo of 0.5 for the Venusian orbit. The cumulative probability for the timing of surface temperatures above 273 K as a function of oribtal distance (c), and the probability of that timing occuring a specific time before present (d).}
\label{fig:Doutputs}
\end{figure}

Placing our Earth-like models in a Martian orbit resulted in mean surface temperatures below freezing for all internal cooling efficiencies (Figure \ref{fig:MarsTs}). Tectonic fluctuations allow the warmest mean path plotted to experience an early and late evolution phase associated with temperatures fluctuating above and below freezing.  For a Venusian orbit, the reference cases had a mean temperatures above freezing from the start. However, increasing the planetary albedo to 0.5 lead to distributions akin to those calculated at the Earth's orbit with a lower albedo. The results from all three orbits indicate that the timing of habitable zone entry depends on how solar energy and internal planetary cooling co-evolve. The two factors interact through a silicate-weathering feedback, whose strength increases as temperature increases. As a result variations in tectonic cooling efficiency affect atmospheric $CO_2$ more so at relatively lower temperatures. Validated thermal paths spanned a wider range of convective velocities early in Earth's history, which further increased the variability in atmospheric $CO_2$. In tandem, these two effects generated a larger potentiality space of surface temperatures earlier in the evolution of model planets. As this occurred around the temperature threshold for liquid water, for Earth orbit and albedo models, the variance due to tectonic cooling efficiencies impacted entry time into the classic habitable zone. The same was true for Earth-like cases in a Venusian orbit but with higher global albedo. For the Martian-orbit suite, the enhanced role of tectonic cooling efficiency increased the variance in surface temperatures but the mean for all paths remained below the freezing point of water. 

We expanded our analysis to cover a range of orbital distances and calculated the cumulative probability of habitable zone entry time for the full range (Figure \ref{fig:DCDF}). We recovered the canonical result that planets further from a star take longer to allow for liquid surface water. However, we also found that variability in the timing of when this happens increases with orbital distance. For example, a distribution of Earth-like planets orbiting at 0.9 AU will be warm enough for liquid water no later than two billion years after formation while an otherwise equivalent distribution orbiting at 1.05 AU will have cases that allow for liquid water close to formation and cases that have entered the liquid water zone in the last hundreds of millions of years. A principal conclusion is that, for orbital distances near that of Earth, the co-evolution of solar energy and internal cooling efficiency determines when a planet will have conditions that allow for  liquid water at its surface. This feeds into the probability that liquid water has existed at the surface of Earth-like planets for a given amount of time (Figure \ref{fig:DEHZ}. Planets akin to the Earth and of the same absolute age could have had habitable surface conditions over time windows that vary by billions of years due to variations in tectonic cooling efficiencies. 

Classic habitable zone modeling has focused on the mean surface temperature path taken by Earth-like planets. However, surface temperatures will fluctuate with some amplitude about the mean based on changes in orbital, solar or interior processes. If the difference between the mean surface temperature and the freezing temperature of water is smaller than this amplitude, then planets could move in and out of the classic habitable zone multiple times over billions of years. From Figure \ref{fig:Earth_Ts}, we see that tectonic fluctuations could place many Earth-like planets in an oscillatory phase for the majority of their lifetime, ceasing within the last billion years. Planets on a cooler mean path could oscillate into habitable conditions; Planets on a warmer mean path could have excursions into and out of snowball states. The pattern of decreasing snowball excursion toward the present day is consistent with historical constraints from the Earth \citep{Kirschvink1992, Kirschvink2000}. The variability of mean paths for our Earth-like planet distribution, when combined with tectonic fluctuations, allows for a range of temperature trajectories that oscillate about global temperature near the freezing point of water. How this may or may not effect planetary life is unclear at this stage but is has been argued that surface temperature oscillation on Earth have fed into evolution and biodiversity \citep[][]{Maruyama2008}. 

We focused on plate-tectonic models and planets of the same size and composition as Earth. Our approach could be expanded to consider different tectonic regimes, planetary compositions and structure. Different weathering parameterizations and continental growth curves could also be considered along with different climate models. However, our principal conclusions is that even for a distribution of similar planets, large variations can exist in terms of when habitable surface conditions could be established. Allowing for other variables would alter absolute model values but variable timing of entry into the classic habitable zone could remain robust, particularly given the added uncertainties that would need to be considered for an expanded model parameter space.

We have also focused on how tectonics influences the temporal establishment of habitable conditions. In terms of the timing at which habitable conditions could no longer be maintained, our models have assumed that $CO_2$ only offsets increasing solar luminosity through a weathering feedback. This reduces the atmospheric $CO_2$ over time, but once levels drop too low, that feedback can not offset increasing solar luminosity. Thus, longer term habitability would require factors not directly included in our models. One potential is dynamic variations of planetary albedo. \citet{Goldblatt2021} found that atmospheric levels of $CO_2$ influence cloud formation inversely. As atmospheric $CO_2$ decreases, cloud cover and planetary albedo increase. This effect could offset increasing solar luminosity and delay an exit from the habitable zone. As atmospheric $CO_2$ levels can remain a function of tectonic efficiency over our full reference model time paths, this suggests that interior processes could modulate both the temporal entrance and exit from the habitable zone (future work that parameterizes albedo with $CO_2$ can lend support to this idea).

The habitable zone is a model that has, from the time of the Kepler mission, guided target selection for exoplanet observations, and it is continuing to do so. The classic model delineates a spatial region around a star where the surface conditions are favorable for liquid water. Temporal considerations have been connected to solar energy (time variable luminosity). A principal conclusion of this study is that internal energy can play as significant a temporal role. Different internal cooling histories can cause billion year variations in the time over which planets of the same absolute age, and sharing several Earth characteristics, could have resided within the liquid water habitable zone. For targeting Earth-like planets, a search strategy favored for practical/pragmatic reasons \citep[][]{Wright2019}, our models provide probability distributions for the time window over which such planets could have maintained habitable conditions (a remote observable is a $P_{CO_2}$ distribution for planets of the same absolute age). Using future observations to confirm or refute those predictions requires multiple targets but is not outside of practical limits \citep[][]{Bean2017,Checlair2019,Lenardic2018c}.  It could provide a step towards confirming the idea that solid planet internal evolution has a significant effect on the establishment of temperate surface conditions and, by association, the potential for surface life \citep[e.g.,][]{Foley2017,Tosi2017}.

As well as guiding target selection, the habitable zone concept can affect how future observations are interpreted and put into context. It is often noted that the early Earth can be viewed, in effect, as an exoplanet. The rationale is that life forms on the early Earth were different than those of the modern Earth as were environmental conditions, most notably atmospheric composition. The search for life beyond our solar system over the next decade will rely on the idea that life can influence atmospheric composition in a way that can create remotely observable biosignatures \citep[][]{Walker2018}. How the evolution of life has played out on the Earth and whether it will or will not be similar on other planets remains debated. However, something that we would argue is less debatable is that evolutionary considerations are more connected to the time window over which a planet has had life than they are to the absolute age of a planet. A second principal conclusion of this study is that tectonic variations can lead to significant differences between the absolute age of a planet and its evolutionary age (the time over which conditions favorable for life have existed). The prediction that follows is that a distribution of planets, sharing multiple Earth characteristics and of the same absolute age, will have a biosignature distribution that spans that of the early and the modern Earth. In addition, finding some planets in that distribution that have no biosignatures does not invalidate the concept of the habitable zone, provided that the model is expanded to include temporal variations associated with internal processes (the main theme of this paper). 

In conclusion, both solar and internal planetary energy influence when surface conditions favorable for liquid water can be established on a terrestrial planet. The influence of internal energy manifests itself via cycling of volatiles between a planets liquid and gaseous outer envelopes and its solid, rocky interior. Volatile cycling depend on the interior cooling efficiency of plate tectonics. Considering a range of viable efficiencies, coupled to surface processes and climate models, leads to the conclusion that the time at which temperate surface conditions are established can vary by billions of years for a distribution of Earth-like planets. The practical implication, for using the habitable zone concept as both a guide for exoplanet target selection and for interpretation of forthcoming observations, is that it should be viewed as both a spatial and a temporal zone, with temporal aspects (e.g. timing of habitability) being a joint function of solar and planetary interior energetic evolutions. 

\section{Appendix}
The modules that constitute our model draw elements from \citet{Walker1981a, Foley2016, Seales2020, Mello2020}. The thermal history module serves as a foundation and we provide a theoretical formulation, discussion of parameter value choices and how we quantified uncertainty for this module. We compared outputs of this module to paleo and present day constraints from the Earth, which served as a filter for identifying Earth-like scenarios. Successful outputs were passed into the remaining modules. These are also detailed below, working from the interior to the surface to the solar forcing module. 

\subsection{Thermal History Model}
Thermal history models have a common underpinning: The Earth's mantle temperature evolves over time based on the balance between heat produced within ($H$) and lost from ($Q$) the mantle according to

\begin{equation}\label{basic}
    C\dot{T_p}=H-Q.
\end{equation}

where $T_p$ is the mantle potential temperature. The radiogenic decay of $^{238}U$, $^{235}U$, $^{232}Th$ and $^{40}K$ produces heat within the mantle according to 
\begin{equation}\label{H}
    H(t)=H_o\sum_{n=1}^4h_nexp(\lambda_nt),h_n=\frac{c_np_n}{\sum_nc_np_n}
\end{equation}
where $H_o$ is a reference heat production. For a given isotope ($n$),  $\lambda_n$, $h_n$, $p_n$ and $c_n$ are the decay constant, the relative contribution to the amount of heat produced, the heat production rate and the concentration of that isotope. Time (\emph{t}) is in billions of years. We calculate relative isotopic concentrations by assuming present day proportions of $U:Th:K=1:4:(1.27x10^4$ and normalizing by total U \citep{Turcotte2002}. The values used in equation \ref{H} are listed in Table \ref{table:radiogenics}.

\begin{table}[h!]
\caption{Radiogenic Heat Production}
\centering
\begin{tabular}{ c c c c c }
\hline\hline
Isotope & $p_n$ $(W/kg)$ & $c_n$ & $h_n$ & $\lambda_n$ $(1/Ga)$ \\
\hline
$^{238}U$  & $9.37\times 10^{-5}$ & 0.9927 & 0.372  & 0.155 \\
$^{235}U$  & $5.69\times 10^{-4}$ & 0.0072 & 0.0164 & 0.985 \\
$^{232}Th$ & $2.69\times 10^{-5}$ & 4.0    & 0.430  & 0.0495 \\
$^{40}K$   & $2.79\times 10^{-5}$ & 1.6256 & 0.181  & 0.555 \\
\hline
\end{tabular}
\label{table:radiogenics}
\end{table}

Thermal convection cools a planetary interior. This is parameterized using the Nusselt-Rayleigh scaling law \citep{Turcotte2002} given by
\begin{equation}\label{Nu_Ra}
    Nu\sim Ra^\beta.
\end{equation}
The Nusselt number \emph{Nu} is a nondimensional heat flux; It is the ratio of convective ($q_{conv}$) to conductive ($q_{cond}$) heat flow across the convecting layer. By Fourier's Law, conductive heat flux is given by: $q_{cond}=\frac{k\Delta T}{D}$, where \emph{k}, \emph{D} and $\Delta T$ are the thermal conductivity, convecting layer thickness and temperature drop across the convecting layer. The Rayleigh number \emph{Ra} is a nondimensional measure of convective vigor and is defined as
\begin{equation}\label{Ra}
    Ra=\frac{\rho g\alpha\Delta TD^3}{\kappa\eta}
\end{equation}
where $\rho$, $\alpha$, $\kappa$, $\eta$ and \emph{g}, are the mantle density, thermal expansivity, thermal diffusivity, viscosity and the acceleration due to gravity. The driving temperature ($\Delta T$) is the temperature between the surface $T_s$ and the interior ($T_p$). Assuming $T_s$ is zero, $\Delta T$ reduces to $T_p$. 

Adding a constant \emph{a} completes Equation \ref{Nu_Ra}. The geometry of the convecting system and the average aspect ratio of its cells influence the value of \emph{a}. Laboratory experiments, boundary layer theory, and numerical experiments can provide estimates of $a$ \citep{Davies1980, Schubert1980}. Alternatively, we can set \emph{a} by scaling the heat flux to its present day value of $Q_o$ and employing a scaling temperature $T_o$ \citep{Christensen1985,Korenaga2003} such that heat flow scales as
\begin{equation}\label{HeatFlow}
    Q = Q_o\left(\frac{T_p}{T_o}\right)^{1+\beta}\left(\frac{\eta (T_o)}{\eta (T_p)}\right)^{\beta}.
\end{equation}
Mantle viscosity ($\eta(T_p)$) is defined as
\begin{equation}\label{eta}
    \eta (T_p)=\eta_o\exp\left(\frac{A_e}{RT_p}\right)
\end{equation}
where  $A_e$, \emph{R} and $\eta_o$ are the activation energy, universal gas constant and scaling constant \citep{Karato771}. For comparison to previous studies, we set $\eta_o$ so that the upper mantle has a viscosity of $10^{19}$ Pa s$\cdot$ s at 1350 $^oC$. Combining equations \ref{Nu_Ra}-\ref{eta} and using the definition of \emph{Nu} leads to the governing equation 
\begin{equation} \label{Ebal}
C\dot{T_p}=H_o\sum_{n=1}^{4} h_nexp\left(-\lambda_nt\right)-Q_o\left(\frac{T_p}{T_o}\right)^{1+\beta}\left(\frac{\eta (T_o)}{\eta (T_p)}\right)^{\beta}.
\end{equation}

Choosing a value for $\beta$ in equation \ref{Ebal} involves making assumptions/hypotheses regarding the dynamics of plate tectonics (\ref{fig:Cartoon}). One hypothesis is that the mantle convects vigorously and that mantle viscosity primarily resists convective motion \citep{Tozer1972b}. Adopting this hypothesis, the earliest thermal history models used a value of 0.33 \citep{Schubert1980,Spohn1982,Jackson1984}. For levels of convection pertinent to the Earth, the scaling exponent is slightly lower, ranging from $0.30<=\beta<=0.32$ \citep{Schubert1985,Lenardic2003}. Models that incorporated analogues to tectonic plates showed that values nearly matching this scaling would be recovered, provided that very weak plate boundaries were also incorporated \citep{Gurnis1989}. Allowing weak plate boundaries to develop dynamically leads to a scaling exponent of 0.29 \citep{Moresi1998}. If plate boundaries are not weak, energy dissipation along them cannot be neglected. If plate strength offers significant resistance, it cannot be neglected either. Both effects lower $\beta$ to between 0 and 0.15 \citep{Christensen1985,Giannandrea1993,Conrad1999b,Conrad1999}. A low viscosity channel below plates - the Earth's asthenosphere \citep{Richards2018} - allows different size plates to have different balances between plate driving and resisting forces \citep{Hoink2011}. Adopting this hypothesis calls for a mixed mode scaling. Considering the distribution of current tectonic plate sizes as a guide, this leads to a global heat flow scaling exponent of $0.15<=\beta<=0.25$ \citep{Hoink2013}. An argument for $\beta<0$ has also been made \citep {Korenaga2003}. The physical basis for this last class of models is that at hotter mantle temperatures enhanced melting would generate a thicker dehydrated layer below oceanic crust. This layer would be responsible for the bulk of plate strength. By this reasoning, hotter mantle temperatures in Earth's past would allow for a thicker, stronger plates, which would slow plate velocities and decrease the rate at which the mantle cooled. 

We account for all the different hypotheses above. This introduces model selection uncertainty into our analysis. To account for this, we will assume the different models historically put forth are unique; however, we will allow for $\beta$ values between them to represent gradational changes between the different hypotheses. Specifically, we tested a range of models with $\beta$ values between -0.15 and 0.3 at intervals of 0.025. For each $\beta$ model, we evaluated combined initial condition and parametric uncertainty (values for each are listed in Table \ref{table:convectiion_parameters}). Initial condition uncertainty for thermal history models comes from uncertainties about post-magma-ocean planetary temperatures. Parametric uncertainty for thermal history models is connected to the values used for radiogenic heating, the heat flow scaling constant, and the scaling temperature. The strength of temperature dependent viscosity is also a model parameter that can be subjected to a range of values. For simplicity, we will not consider that explicitly herein, as it is connected to variations in the mantle Rayleigh number, \emph{Ra}, which will already be subjected to a range of variations due to the variations in the other parameters noted. For each unique combination of $\beta$, choice of initial condition and choice of parameters we also calculated the uncertainty due to un-modeled factors using a perturbed physics approach \citep{Astrom2008,Seales2019}. This provided a measure of structural uncertainty for models \citep{Guckenheimer1983,Kennedy2001}. The result of this analysis was an output ensemble (see \citet{Seales2019} for a full description of this method). Including all sources of uncertainty, our analysis involved computing more than 1.25 million model evolutions.

The success or failure of any model ensemble was determined by comparing the ensemble mean and uncertainty bounds to paleo and present day constraints. For a full description of our analysis see \citet{Seales2020}. We will briefly summarize here. For paleo constraints we use the results of \citet{Ganne2017} who calculated uncertainty bounds on mantle temperatures over time. For our present day temperature constraint we used a value of 1350 $^oC$ $\pm 50$ $^oC$ \citep{Herzberg2008}. Our second present day constraint was the mantle Urey ratio, \emph{Ur}, which is the the ratio of \emph{H} to \emph{Q}. \citet{Jaupart2007} estimate it to be between 0.3 and 0.5. Allowing for continents, the \emph{Ur} upper bound can be extended \citep{Grigne2001, Lenardic2011}. \citet{Lenardic2011} showed that, for present day continental land fractions, heat flows are consistent for mantles with and without continental coverage. Therefore, the upper \emph{Ur} bound, for models that do not directly include continental effects, can be extended to approximately 0.7 as an adjustment to account for present day continental effects. If an ensemble matches all data constraints, within uncertainties, we proceed with the remainder of our analysis as detailed below.

\begin{table}[h]
%%\begin{adjustwidth}{-.75in}{-.75in}  
\caption{Model Parameters}
\centering
\begin{tabular}{ c c c c }
\hline\hline
Parameter & Values & Units & Description \\
\hline
$T_i$ & 1000, 1250, 1500, 1750, 2000 & $^oC$ & Initial Temperature \\
$T_o$ & 1300, 1350, 1400 & $^oC$ & Scaling Temperature \\
$Q_o$ & 3.0e13, 3.5e13, 4.0e13 & TW & Scaling Heat Flow \\
$H_o$ & 2.19e13, 2.55e13, 2.92e13, 4.38e13, 5.12e13, & TW & Initial Radiogenics \\
& 5.84e13, 6.57e13, 7.66e13, 8.76e13, 1.02e14, & & \\
& 1.09e14, 1.17e14, 1.28e14, 1.46e14 & & \\
$\eta_o$ & 2.21e9 & Pa$\cdot$s & Viscosity constant \\
A & 300 & kJ mol$^{-1}$ & Activation Energy \\
R & 8.314 & J / (mol$\cdot$K) & Universal Gas Constant \\
\hline
\end{tabular}
\label{table:convectiion_parameters}
%%\end{adjustwidth}  
\end{table}

Convective mantle overturn provides a key link between interior and surface processes via volatile cycling. Following \citet{McGovern1989, Franck1998} we connect the interior module to volatile cycling through an areal spreading rate for tectonic plates ($U_c$) according to  
\begin{equation}\label{convective_vel}
U_c=\frac{(\gamma q_{conv})^2\pi\kappa A_{ob}}{4k^2(T_p)^2}
\end{equation}
where $A_{ob}$ is the fractional area of ocean basins. The scalar $\gamma$ takes into account variations in heat flow due to plate thickening. Thinner plates nearer the mid-ocean ridges allow for higher heat flows, whereas colder, thicker plates near trenches allow for reduced heat flows. 

The fraction of surface area taken up by continents ($A_{cc}$) grows with time, reducing $A_{ob}$ closer to present day model times. Continental growth remains debated and the growth rate varies greatly between studies. We adopt the growth curve of \citet{Rosing2010}, which assumes sigmoidal growth that correlates continental volume with continental surface area according to 
\begin{equation}\label{cont_growth}
    A_{cc}=\frac{1}{1+exp[-2(t-2)]}
\end{equation}
\begin{equation}
    A_{ob}=1-f_{cc}A_{cc}
\end{equation}
\begin{equation}
    A_{ob}=A_EA_{ob}
\end{equation}
where $f_{cc}$ is the present day fraction of Earth's surface area ($A_E$) covered by continents. In this model, most of Earth's crust formed between 3.0 and 1.5 Ga. Different growth curves could be adopted but this would enhance model parameter space, and associated uncertainty. As with other potential model expansions this will alter absolute model values but is less likely to alter our principal conclusion that billion year variations can exist for the temporal establishment of habitable conditions on a distribution of planets that share multiple common factors (size, mass, orbital distance, tectonic mode). 

\subsection{Water Cycling Model}
The dominant sources and sinks in the water cycle are degassing at mid-ocean ridges and regassing via subducting plates. We follow the model described in \citet{Cowan2014a}. Warm mantle rises beneath mid-ocean ridges and begins melting at a depth ($d_m$). The melt contains water, a fraction ($f_d$) of which is lost to the surface reservoir according to 
\begin{equation}
    f_d=min\left[f_{d,E}\left(\frac{P}{P_E}\right)^{-\mu},1\right]
\end{equation}
\begin{equation}
    P=\rho_wgd_{oc}
\end{equation}
\begin{equation}
    d_{oc}=\frac{M_{ow}}{\rho_wA_{ob}}
\end{equation}
where $f_{d,E}$ is the nominal degree of melting, $P$ is the pressure of the overlying water, $P_E$ is the present day value of pressure and $\mu$ describes the pressure dependence. As the mass of ocean water ($M_{ow}$) at Earth's surface has varied over its history, so too does the depth of the overlying water column, $d_{ob}$. Water density ($\rho_w$) is a constant. The mass flow rate of water degassed from the mantle ($\left[\frac{dM_{w}}{dt}\right]$) is
\begin{equation}\label{Mowd}
    \left[\frac{dM_{w}}{dt}\right]_d=\rho_{ow}d_mf_dU_c.
\end{equation}

Surface water moves along fractures within the plate, hydrothermally altering the rocks. This physically transfers water from the surface reservoir into the plate. The depth to which hydrothermal alteration occurs ($d_h$) varies. At first order, we assume these fractures penetrate the thickness of the basaltic layer ($d_b$) or some pressure-scaled depth, whichever is thinner, according to 
\begin{equation}
    d_h=min\left[d_{h,E}\left(\frac{P}{P_E}\right)^\sigma ,d_b\right].
\end{equation}
The parameter $d_{h,E}$ is the nominal hydration depth. We assume $d_b$ to be a constant. The subducting plate carries the bound water back into the mantle. Although heating of the downwelling plate releases some of the water back to the surface through arc volcanism, a fraction ($f_r$) returns to the mantle. Using this same model, \citet{Mello2020} found a value of 0.4 for $f_r$. For a more complex water cycling model that used Earth data as constraints, \citet{Seales2020a} found $f_r$ of 0.1, so this should produce the correct first order behavior within this module. The mass flow rate of water regassed into the mantle is ($\left[\frac{dM_{w}}{dt}\right]$) is
\begin{equation}\label{Mowr}
    \left[\frac{dM_{w}}{dt}\right]_r=f_{bas}\rho_{bas}d_hf_rU_c
\end{equation}
where $f_{bas}$ is the mean mass fraction of water in the basaltic layer. We allow the mass of surface water ($M_{sw}$) to evolve self-consistently according to
\begin{equation}
    \frac{dM_{sw}}{dt}=\left[\frac{dM_{w}}{dt}\right]_d-\left[\frac{dM_{w}}{dt}\right]_r.
\end{equation}

\subsection{Weathering Function}
We assume that silicate-weathering determines the amount of atmospheric $CO_2$. We neglect the effects of seafloor weathering to provide a first order analysis. The model of \citet{Walker1981a} tracks the kinetics of weathering ($W$), which depend on the presence of greenhouse gases and surface temperature, according to 
\begin{equation}\label{W}
    \frac{W}{W_o}=\left(\frac{P_{H_2O}}{P_{H_2O}^o}\right)^a\left(\frac{P_{CO_2}}{P_{CO_2}^o}\right)^bexp\left[-\frac{E_a}{R}\left(\frac{1}{T_s}-\frac{1}{T_{sat}}\right)\right]
\end{equation}
where the script $o$ represents present day values, except for $CO_2$, which we set to pre-industrial levels. We performed a structural stability analysis by randomly perturbing the parameters $a$ and $b$. We found the model to be stable and with levels of structural uncertainty less than a few percent. The constant $E_a$ is the activation energy of the weathering reaction and $T_{sat}$ is the saturation temperature. In this model, weathering increases with increased $P_{CO_2}$, surface temperature ($T_s$) and water saturation pressure ($P_{H_2O}$). For the latter, higher surface temperatures lead to more rain and increased weathering rates. 

To approximate the first order effects of different tectonic cooling efficiencies on surface temperature, we follow \citet{Walker1981a, Mello2020} in assuming that weathering and $CO_2$ outgassing are proportionate. Outgassing scales with $S_r$, so we equate $W$ and $S_r$ in Equation \ref{W} and rearrange to find $P_{CO_2}$:
\begin{equation}\label{PCO2}
    P_{CO_2}=P_{CO_2}^o\bigg\{\left(\frac{U_c}{U_c^o}\right)\left(\frac{P_{H_2O}}{P_{H_2O}^o}\right)^{-a}\times exp\left[\frac{E_a}{R}\left(\frac{1}{T_s}-\frac{1}{T_{sat}}\right)\right]\bigg\}^\frac{1}{b}
\end{equation}
Our assumption of equilibrium between these reservoirs remains valid for time steps larger than $10^5$ years, the time it takes for the two to equilibrate. This model would break down if all weatherable rock vanished. More complex models follow the fluxes of carbon between the mantle and surface reservoirs \citep{Foley2016}. As with other potential model expansions this will alter absolute model values but is less likely to alter our principal conclusion that billion year variations can exist for the temporal establishment of habitable conditions as carbon cycling in those types of models will also depend on tectonic cooling efficiencies. 

\subsection{Climate Model}
We follow \citet{Foley2016} in relating surface temperature and $CO_2$ according to the model of \citet{Walker1981a}: 
\begin{equation}\label{surftemp}
    T_s=T_{s,o}+2(T_e-T_{e,o})+4.6\left(\frac{P_{CO_2}}{P_{CO_2}^o}\right)^{0.346}-4.6
\end{equation}
where $T_e$ is the effective temperature and the $o$ superscripts represent present day values. Parameterized this way, the model assumes a water saturated atmosphere. The effective temperature relates to solar radiation according to
\begin{equation}
    T_e=\left[\frac{S(1-A)}{4\sigma}\right]^\frac{1}{4}
\end{equation}
where $S$ is the solar irradiance, $A$ is albedo and $\sigma$ is the Stefan-Boltzmann constant. $A$ remains fixed for all time at the stated value. This model approximates radiative-convective models to first order \citep{Kasting1986}. Our purpose pertains to the habitable zone entry time, which we define as when surface temperatures reach 273 K, so we do not concern ourselves with effects due to a moist or runaway greenhouse that would occur at higher temperatures. 

We allow $S$ to vary in our model based on a solar luminosity model present in \citet{Mello2020}:
\begin{equation}\label{L}
\begin{split}
L(t)& \\
& = 1.0424 \times \left(\right.-2.245 + 0.7376 \times t + 16.03 \\
& - 0.2348 \times t^2 - 4.596 \times t - 44.2 \\
& + 0.1212 \times t^2 +10.5 \times t + 59.23 \\
& - 0.2047 \times t^2 - 10.43 \times t - 38.59 \\
& + 0.1132 \times t^2 + 3.82 + 10.46 \left. \right)
\end{split}
\end{equation}
where $t$ is given in billions of years. They used the main sequence stellar luminosity parameterized by \citet{Rushby2013}, considering the data of \citet{Baraffe1998}, and introduced a small correction to the model such that the solar luminosity matched that of our own solar system at 4.57 billion years. To obtain the solar flux at Earth, we divide $L$ by the surface area of a sphere with one astronomical unit, obtaining a present-day value of $S_o=1368$ $Wm^2$. 

\begin{table}[h]
\caption{Model constants, scaling values and parameter ranges}
\centering
\begin{tabular}{ c c c c c }
\hline\hline
Module & Symbol & Parameter & Value & Units    \\
\hline \\
$H_o$ & Initial radiogenic concentration & $1.25\times 10^{-7}$ & $Wm^{-3}$  \\
$\lambda$ & Decay constant & 0.34  & $Gyr^{-1}$ \\
$\alpha$ & Thermal expansivity & 2e-5 & $K^{-1}$ \\
$\kappa$ & Thermal diffusivity & 1e-6 & $m^2s^{-1}$ \\
$T_s$ & Surface Temperature  & 273 & $K$ \\
$\eta_{ref}$ & Reference viscosity & $1e21 $ & $Pa*s$ \\
$A_e$ & Activation energy & 3e5 & $Jmol^{-1}$ \\
$R$ & Universal gas constant & 8.314 & $J(K*mol)^{-1}$ \\
$T_{ref}$ & Reference temperature & 1855 & $K$ \\
$Ra_c$ & Critical Rayleigh number & 1100 & - \\
$c_p$ & Heat capacitance & 1400 & $J(kg*K)^{-1}$ \\
$k$ & Thermal conductivity & 4.2 & $W(m\times K)^{-1}$ \\
$T_o$ & Scaling temperature & 1600 & $K$ \\
$q_o$ & Scaling convective heat flow & 0.069 & $Wm^{-2}$ \\
$\eta_o$ & Scaling viscosity & 4.45e19 & $Pa*s$ \\
$M_\oplus$ & Mass of Earth & 5.97e24 & $kg$ \\
$R_\oplus$ & Radius of Earth & 6371 & $km$ \\
$G$ & Gravitational constant & 6.67408e-11 & $Nm^2kg^{-2}$ \\
$\beta$ & Tectonic cooling efficiency constant & 0-0.33 & - \\
$M_p$ & Mass of Planet & 0.1-5 $M_\oplus$ & - \\
$R_p$ & Planet radius & Calculated & $km$ \\
$R_c$ & Core radius & Calculated & $km$ \\
$\rho$ & Mantle density & Calculated  & $kgm^{-3}$ \\
$g$ & Surface gravity & Calculated & $ms^{-2}$ \\
\hline
\end{tabular}
\label{table:convectiion_parameters2}
\end{table}

\bibliography{EHZ_Refs}

\end{document}